\documentclass[12pt]{article}
\usepackage{amsfonts}
\usepackage{amsmath}
\usepackage{amssymb}
\usepackage{amsthm}
\usepackage{accents}
\usepackage{youngtab}
\usepackage{braket}
\usepackage{graphicx}
\usepackage[svgnames]{xcolor}
\usepackage{ytableau}
\usepackage{here}
\usepackage{comment}
\usepackage{tikz}
\usepackage{cite}
\usepackage{feynmf}

\newcommand{\ba}{\begin{eqnarray}}
\newcommand{\ea}{\end{eqnarray}}
\newcommand{\nn}{\nonumber}

\newcommand{\hg}{\hat{\gamma}}

\newcommand{\tyng}{\tiny\yng}

\newcommand{\cN}{\mathcal{N}}
\newcommand{\cW}{\mathcal{W}}

\newcommand{\cP}{\mathcal{P}}
\newcommand{\lt}{\left}
\newcommand{\rt}{\right}
\newcommand{\cO}{\mathcal{O}}

\makeatletter
\newcommand{\douwidehat}[2]{%
  \sbox0{$\m@th#1\widehat{\hphantom{#2}}$}%
  \sbox2{$\m@th#1x$}
  \sbox4{$\m@th#1#2$}
  \dimen0=\ht0
  \advance\dimen0 -.8\ht2
  \dimen2=\dp4
  \rlap{%
    \raisebox{\dimexpr\dimen0-\dimen2}{%
      \scalebox{1}[-1]{\box0}%
    }%
  }%
  {#2}%
}
\makeatother

\usepackage[colorlinks=true,linkcolor=black,citecolor=black,urlcolor=black]{hyperref}

\begin{document}

\unitlength=1mm
\begin{titlepage}
\begin{flushright}
UT-17-18
\end{flushright}
\vskip 12mm
\begin{center}
  {\LARGE {\bf Maulik-Okounkov's R-matrix from Ding-Iohara-Miki algebra}}
\end{center}
\vskip 2cm
\begin{center}
  {M. Fukuda, K. Harada, Y. Matsuo and R.-D. Zhu}
\end{center}
\vskip 2cm
\begin{center}
{\it 
Department of Physics, the University of Tokyo\\
Hongo 7-3-1, Bunkyo-ku, Tokyo 113-0033, Japan}
\end{center}
\vfill
\begin{abstract}
The integrability of 4d $\cN=2$ gauge theories has been explored in various contexts, for example the Seiberg-Witten curve and its quantization. Recently, Maulik and Okounkov proposed that an integrable lattice model is associated with the gauge theory, through an R-matrix, to which we refer as MO's R-matrix in this paper, constructed in the instanton moduli space. In this paper, we study the R-matrix using the Ding-Iohara-Miki (DIM) algebra. We provide a concrete boson realization of the universal R-matrix in DIM and show that the defining conditions for MO's R-matrix can be derived from this free boson oscillator expression. Several consistency checks for the oscillator expression are also performed. 
\end{abstract}
\vfill
\end{titlepage}

\section{Introduction}

It has been more than 30 years ago after the discovery of the integrable nature of 4d $\cN=2$ super Yang-Mills theories \cite{Donagi:1995cf,Itoyama:1995nv, Nakatsu:1995bz,SW-Integrable}. 
After intensive studies, it turned out that at least for a rather large family of 4d $\cN=2$ theories, the Class ${\cal S}$ theories \cite{Gaiotto09}, each gauge theory in the Coulomb branch is associated to a Hitchin integrable system \cite{GMN}.
The quantization of such an integrable model is described by the introduction of $\Omega$-background \cite{Nekrasov2004} which has two deformation parameters.  When one parameter is tuned to be zero, which is referred to as Nekrasov-Shatashvili limit, the instanton partition functions have new connections with integrable quantum mechanical systems such as Toda-chain, Calogero system and Bethe ansatz equation \cite{Nekrasov:2009rc}.
For the general choice of the parameters, such integrable structure is promoted to quantum field theoretical one described by qq-character which corresponds to infinite dimensional quantum algebras \cite{NPS}.  The explicit algebraic realization was demonstrated by \cite{BMZ,Kimura-Pestun,Nekrasov:2015wsu,5dBMZ} where the relevant symmetries were identified with (quantum) W-algebra \cite{Shiraishi:1995rp}, Ding-Iohara-Miki (DIM) algebra \cite{Ding-Iohara,Miki} and SH$^c$ algebra for a degenerate limit \cite{SHc,Tsymbaliuk14,Prochazka}. 

A more direct link with the quantum integrability structure with such systems was explored in \cite{MO} where an R-matrix was constructed in the cohomology ring of the instanton moduli space. We will refer to it as Maulik-Okounkov's (MO's) R-matrix in this paper. From the construction of the R-matrix, the SH$^c$ algebra was identified as the $\widehat{\mathfrak{gl}_1}$ Yangian. 
A remarkable property of MO's R-matrix is that it acts on the tensor product of two Fock spaces instead of the finite dimensional spin degrees of freedom.  By composing several Fock spaces with R-matrices, we can realize the Hilbert space of $\mathfrak{u}(1)\times\cW_N$ algebras as a Yangian.  It was essential in \cite{SHc} to prove the AGT conjecture \cite{AGT,Wyllard}. While the appearances of the algebras are totally different, the correspondence between the representation is exact even for the degenerate representations \cite{FNMZ}.

To push forward in the direction of understanding these integrable models and especially the physical meaning of MO's R-matrix in gauge theories, we consider a generalization to the 5d version of AGT \cite{Awata-Yamada}.
The compactification on $S^1$ circle implies that the corresponding $\cW$-algebra should be $q$-deformed. 
As the representations of W-algebra is universally covered by SH$^c$ (or Yangian), the q-deformed version is described by DIM \cite{FHSSY}, and the 5d AGT relation can be proved with its help (see, for example, \cite{5dBMZ}). 
As its 4d sibling, the DIM algebra also possesses an R-matrix \cite{DIMBethe}. 
This fact can be understood from the aspect that DIM is a quantum-group analogy for the Yangian, and thus is equipped with a universal R-matrix \cite{RTF90}. 
The matrix elements of the R-matrix at the first several levels were computed in \cite{MNagoya,D-Ohkubo}. 

From the viewpoint of string theory, a merit to consider the q-deformed version is that the duality structure is more manifest.
It is known that the DIM has an SL(2,$\mathbb{Z}$) automorphism \cite{Miki}.  Since it interchanges the topological string amplitudes including D and NS branes through the topological vertex \cite{aganagic2005topological,iqbal2009refined} realized as an intertwinner of DIM \cite{Awata2011}, it may be identified with the duality symmetry of type IIB superstring.
Thus the integrability property begins to have a new perspective in the relevance of non-perturbative duality in the brane-web (see for example \cite{Mironov:2016yue,Mironov:2016yue,Awata:2016riz,BFHMZ}). 

The purpose of this paper is to explore an oscillator realization of the R-matrix for DIM.
While there are some works in this subject (see for example,\cite{DIMBethe,feigin2016finite,MNagoya,okounkov2016quantum}), 
the bosonic representation along MO's line is new.
Since MO's R-matrix gives the bosonized Hamiltonians of (generalized) Calogero-Sutherland system, our result gives those for Ruijsenaars-Schneider model \cite{RS-1,RS-2,Konno95,Konno96,RS-review}. These Hamiltonians describes the structure of the cohomology ring for (K-theoretical) moduli space of instanton moduli spaces.

We organize this paper as follows.  In section \ref{s:DIM}, we briefly describe the DIM algebra and its bosonic (horizontal) representation.
In section \ref{s:MO}, we first review the R-matrix proposed by Maulik-Okounkov \cite{MO} and explain how conserved charges can be constructed out of it.  Next, we use the notion of the universal R-matrix to derive the property which should be obeyed by R-matrix for DIM.
In section \ref{Exp-R}, we expand the R-matrix in terms of the spectral parameter and derive the recursion relation among coefficients.  We also give the first two terms in the expansion.
In section \ref{s:charges}, we derive some conserved charges from the R-matrix.
Finally in section \ref{Ohkubo}, we calculate the action of R-matrix on the generalized Macdonald basis and shows the consistency with the previous result \cite{MNagoya,D-Ohkubo}.


\subsection*{Notation Conventions}

Let us explain our convention on notations here. There are two parameters introduced for the $\Omega$-background, $q$ and $t$, but it is more convenient for us to introduce a third one $p$, which is related to $q$ and $t$ by 
\ba
p=\frac{q}{t}.\nn
\ea
There are also several types of bosons used in this article and since it seems to be confusing, we list all of them here. 
\begin{itemize}
\item $b_n$ satisfying $[b_m, b_n]=\frac{1}{m}(1-q^{-m})(1-t^m)(1-p^m)(\hg^m-\hg^{-m})\hg^{-|m|}\delta_{m+n,0}$ ($\hg$ is a central element in the DIM algebra) is a generator in the DIM algebra. We note that in different representations of DIM, it takes different forms, for example in the vertical representation, it is a diagonal operator, and in the horizontal representation, it is mapped to the $q$-boson, $a_n$. 
\item The $q$-boson $a_n$ satisfies the commutation relation $[a_m, a_n] = m\frac{1-q^{|m|}}{1-t^{|m|}}\delta_{m+n,0}$. It is used in the horizontal representation of the DIM algebra. We note that it has no definite coproduct structure. 
\item $\alpha_n$ is the normal boson oscillator defined by $[\alpha_n,\alpha_m]=n\delta_{n+m,0}$. It is only used in the degenerate limit $q\rightarrow 1$. We note that the vertex operator $\alpha(z)$ has nothing to do with $\alpha_n$, especially the latter one does not give the mode expansion of $\alpha(z)$. 
\end{itemize}
We have to work on tensor porudcts of Fock spaces/representation spaces. We use the superscript $^{(i)}$ to denote the operator living in the $i$-th Fock space/representation space. For example, $\alpha^{(2)}_n$ is the boson operator $\alpha_n$ in the second Fock space, and $x^{+(0)}(z)$ is the generator $x^+(z)$ defined in the $0$-th (auxiliary) copy of the DIM algebra. We also use short notations to express some special combination of bosons in several Fock spaces. 
\ba
\alpha^\pm_n:=\frac{1}{\sqrt{2}}\lt(\alpha^{(1)}_n\pm \alpha^{(2)}_n\rt),\quad
a^\pm_n:=\frac{1}{\sqrt{2}}\lt(a^{(1)}_n\pm a^{(2)}_n\rt),\quad
a^{p-}_n:=a_n^{(2)}-p^{\frac{|n|}{2}}a_n^{(1)}.\nn
\ea
For a state $\ket{\psi}$ in the $i$-th representation space, however, we use the notation $\ket{\psi}_i$.

\section{DIM algebra and its universal R-matrix}\label{s:DIM}

In this section, we review the defining relations of the DIM algebra and the universal R-matrix. We adopt the same convention of notations as in our previous paper \cite{BFHMZ}. 

\subsection{DIM algebra}

The DIM algebra, which will be denoted as $U_{q_1,q_2}(\widehat{\mathfrak{gl}_1})$ in this paper, has two parameters $q_1=q$ and $q_2=t^{-1}$ and satisfies the following defining relations.
\ba
&&[\psi^\pm(z),\psi^\pm(w)]=0,\quad \psi^+(z)\psi^-(w)=\dfrac{g(\hg w/z)}{g(\hg^{-1}w/z)}\psi^-(w)\psi^+(z),\label{def-1}\\
&&\psi^+(z)x^\pm(w)=g(\hg^{\mp1/2}w/z)^{\mp1}x^\pm(w)\psi^+(z),\\
&&\psi^-(z)x^\pm(w)=g(\hg^{\mp1/2}z/w)^{\pm1}x^\pm(w)\psi^-(z),\\
&&x^\pm(z)x^\pm(w)=g(z/w)^{\pm1}x^\pm(w)x^\pm(z),\\
&&[x^+(z),x^-(w)]=\dfrac{(1-q_1)(1-q_2)}{(1-q_1q_2)}\left(\delta(\hg^{-1}z/w)\psi^+(\hg^{1/2}w)-\delta(\hg z/w)\psi^-(\hg^{-1/2}w)\right),
\ea
where 
\ba
g(z):=\prod_{i=1,2,3}\dfrac{1-q_i z}{1-q_i^{-1}z},\quad q_3:=(q_1q_2)^{-1},\nn
\ea
$\hat{\gamma}$ is a central element of the algebra, and $\delta(z)=\sum_{n\in\mathbb{Z}}z^n$ is the delta-function. In addition, it also satisfies the Serre relations \cite{FFJMM1}, 
\ba
\lt[x^+_0,\lt[x^+_1,x^+_{-1}\rt]\rt]=0,\quad \lt[x^-_0,\lt[x^-_1,x^-_{-1}\rt]\rt]=0.
\ea
The mode expansion of each generator is given by 
\ba
x^\pm(z)=\sum_{n\in\mathbb{Z}}x^\pm_nz^{-n},\quad \psi^\pm(z)=\sum_{n\geq 0}\psi^\pm_{\pm n}z^{\mp n}.
\ea
These defining relations are a two-parameter generalization of the Drinfeld realization of the quantum group \cite{Drinfeld-realization} and thus the DIM algebra is also sometimes called the quantum toroidal algebra (associated with 
$\mathfrak{gl}_1$).

This algebra has the following known coproduct structure,
\begin{equation}
\begin{split} 
  &\Delta(\hg^{\pm\frac{1}{2}})=\hg^{\pm\frac{1}{2}}\otimes \hg^{\pm\frac{1}{2}},\\
  &\Delta(\psi^{\pm}(z))=\psi^{\pm}(\gamma_{(2)}^{\pm\frac{1}{2}}z)\otimes\psi^{\pm}(\gamma_{(1)}^{\mp\frac{1}{2}}z),\\
  &\Delta(x^+(z))=x^+(z)\otimes1+\psi^-(\gamma_{(1)}^{\frac{1}{2}}z)\otimes x^+(\gamma_{(1)}z),\\
    &\Delta(x^-(z))=x^-(\gamma_{(2)}z)\otimes\psi^+(\gamma_{(2)}^{\frac{1}{2}}z)+1\otimes x^-(z),
    \end{split}
\end{equation} 
where, $\gamma_{(1)}^{\pm\frac{1}{2}}=\hg^{\pm\frac{1}{2}}\otimes1$ and  $\gamma_{(2)}^{\pm\frac{1}{2}}=1\otimes\hg^{\pm\frac{1}{2}}$. 

Given a quantum group, a special operator called R-matrix is usually inherited from its quasi-trangular Hopf algebra nature. This R-matrix, which is often called the universal R-matrix, has the following properties. 
\ba
&&{\bf R}\Delta(e)(z){\bf R}^{-1}=\cP\circ\Delta(e)(z),\ \ {\rm for}\ e\in {\rm quantum\ group}),\\
&&(\Delta\otimes 1){\bf R}={\bf R}_{13}{\bf R}_{23},\quad (1\otimes \Delta){\bf R}={\bf R}_{13}{\bf R}_{12},
\ea
where $\cP$ is the operator which exchanges two copies of algebras (quantum groups), i.e. $\cP(a\otimes b)=b\otimes a$, and the notation ${\bf R}_{ij}$ denotes the R-matrix acting on the $i$-th and $j$-th algebras. Such an R-matrix automatically meets the Yang-Baxter equation. Since the DIM algebra is a generalization to the quantum groups, we expect a similar R-matrix structure in DIM, and indeed in \cite{DIMBethe}, it was argued that the DIM algebra is a quasitrangular Hopf algebra and equipped with the universal R-matrix ${\bf R}(u)$. This object will be the main character of the paper. 

\subsection{Horizontal Representation of DIM Algebra and $q$-deformed $\cW$-algebras}

In addition to $\hat{\gamma}$, there is one more central element $\psi^+_0/\psi^-_0$ in the DIM algebra. The representation of the algebra can thus be parameterized with two numbers $(\ell_1,\ell_2)$ such that $\hat{\gamma}=q_3^{\ell_1/2}$ and $\psi^+_0/\psi^-_0=q_3^{\ell_2}$. A large family of the representations is parameterized by $(\ell_1,\ell_2)\in\mathbb{Z}\times\mathbb{Z}$, among which the SL(2,$\mathbb{Z}$) automorphism of the algebra acts as 
\ba
S\cdot (l_1,l_2)=(-l_2,l_1),\quad T\cdot (l_1,l_2)=(l_1,l_1+l_2).
\ea
We often refer to the transformation $S$, which is a $90^\circ$ rotation on the representation plane, as the S-duality of the algebra. 

The $(1,0)$ representation is called the horizontal representation of the DIM algebra in the literature. It is characterized by the $q$-boson representation.
Let us introduce the following vertex operators: 
\begin{equation}
\begin{split}
  &\eta(z)=:\mathrm{exp}(\sum_{n\neq0}\frac{1-t^{-n}}{n}a_{-n}z^n):,\\
  &\xi(z)=:\mathrm{exp}(-\sum_{n\neq0}\frac{1-t^{-n}}{n}p^{-\frac{|n|}{2}}a_{-n}z^n):,\\
  &\varphi^+(z)=\mathrm{exp}(-\sum_{n>0}\frac{1-t^n}{n}(1-p^{-n})p^{\frac{n}{4}}a_nz^{-n}),\\
  &\varphi^-(z)=\mathrm{exp}(\sum_{n>0}\frac{1-t^{-n}}{n}(1-p^{-n})p^{\frac{n}{4}}a_{-n}z^n),
\end{split}
\end{equation}
where $:\bullet:$ denotes the normal ordering, $p=\frac{q}{t}$, and $a_n$ is the q-deformed boson oscillator satisfying 
\begin{equation}
  [a_m, a_n] = m\frac{1-q^{|m|}}{1-t^{|m|}}\delta_{m+n,0}.
\end{equation}
The horizontal representation is defined by these vertex operators: 
\begin{equation}
  \rho_u(\hg^{\pm\frac{1}{2}})=p^{\mp\frac{1}{4}},\quad \rho_u(\psi^{\pm}(z))=\varphi^{\pm}(z),\quad \rho_u(x^+(z))=u\eta(z),\quad\rho_u(x^-(z))=u^{-1}\xi(z).
\end{equation}
$u\in \mathbb{C}$ is the weight of the representation. 

It is convenient to introduce the boson oscillators $b_n$ as the mode expansion of $\log\psi^\pm$, 
\begin{equation}
 \psi^+(z)=\psi^+_0\exp\lt(\sum_{n>0}b_n\hg^{\frac{n}{2}}z^{-n}\rt),\quad  \psi^-(z)=\psi^-_0\exp\lt(-\sum_{n>0}b_{-n}\hg^{\frac{n}{2}}z^n\rt).
\end{equation}
From the algebraic relation (\ref{def-1}) and the coproduct of $\psi^\pm$, we see that $b_n$ satisfies
\begin{equation}
\begin{split}
  &[b_m, b_n]=\frac{1}{m}(1-q^{-m})(1-t^m)(1-p^m)(\hg^m-\hg^{-m})\hg^{-|m|}\delta_{m+n,0},\\
  &\Delta(b_n)=b_n\otimes\hg^{-|n|}+1\otimes b_n.
\end{split}
\end{equation}
In the horizontal representation, $b_n$ and $\psi_0^{\pm}$ are respectively mapped to 
\begin{equation}
b_n\mapsto -\frac{1-t^n}{|n|}(p^{\frac{|n|}{2}}-p^{-\frac{|n|}{2}})a_n, \quad  \psi^{\pm}_0\mapsto 1.
\end{equation}

It is known \cite{FHSSY} that one can embed the $q$-deformed 
${\cal W}_m$ algebra in the $(m,0)$ representation of the DIM.
$(m,0)$ representation can be constructed by taking the coproduct of $m$ $(1,0)$ representations.
For $m=2$ case, the representation is isomorphic to a direct sum of
$U(1)$ current and q-Virasoro generators.
The latter one is realized as \cite{FHSSY}
\ba
\rho_{u_1,u_2}(\Delta(t(z))):=\rho_{u_1}\otimes\rho_{u_2}(\Delta(t(z)))=:u_1\Lambda_1(z)+u_2\Lambda_2(z),
\ea
where 
\begin{equation}
  t(z)=\alpha(z)x^+(z)\beta(z),
\end{equation}
with 
\begin{equation}
\begin{split}
  &\alpha(z)=\mathrm{exp}(-\sum_{n=1}^{\infty}\frac{1}{\hg^n-\hg^{-n}}b_{-n}z^n),\\
  &\beta(z)=\mathrm{exp}(\sum_{n=1}^{\infty}\frac{1}{\hg^n-\hg^{-n}}b_nz^{-n}).
\end{split}
\end{equation}
For later convenience, we list the expressions of $\alpha$, $\beta$ and $\Lambda_{1,2}$ in the horizontal representation\footnote{{\bf Remark:} It might be confusing that substituting $\Delta(a_n)=p^{|n|/2}a^{(1)}_n+a^{(1)}_n$ into the above expression of $\rho_u(\alpha(z))$ does not give us $\rho_{u_1}\otimes\rho_{u_2}(\Delta(\alpha(z)))$. We note that the $q$-boson does not have a consistent coproduct structure, as can be seen from the fact that $\rho_u(x^+(z))=u\exp(\sum_{n\neq 0}\frac{1-t^{-n}}{n}a_{-n}z^n)$ and $\Delta(x^+(z))=x^+(z)\otimes 1+\psi^-(\gamma_{(1)}^{1/2}z)\otimes x^+(\gamma_{(1)}z)$. The correct formula can only be obtained by first taking the coproduct in the expression in terms of generator of the DIM algebra and then performing the representation map. }. 
\begin{equation}
\begin{split}
  &\rho_u(\alpha(z))=\exp\lt(-\sum_{n=1}^{\infty}\frac{1-t^{-n}}{n}a_{-n}z^n\rt),\\
  &\rho_{u_1}\otimes\rho_{u_2}(\Delta(\alpha(z)))=\exp\lt(-\sum_{n=1}^{\infty}\frac{1-t^{-n}}{n(p^{\frac{n}{2}}+p^{-\frac{n}{2}})}(p^{\frac{n}{2}}a_{-n}^{(1)}+a_{-n}^{(2)})z^n\rt),\\
 &\rho_u(\beta(z))=\exp\lt(\sum_{n=1}^{\infty}\frac{1-t^n}{n}a_nz^{-n}\rt),\\
  &\rho_{u_1}\otimes\rho_{u_2}(\Delta(\beta(z)))=\exp\lt(\sum_{n=1}^{\infty}\frac{1-t^n}{n(p^{\frac{n}{2}}+p^{-\frac{n}{2}})}(p^{\frac{n}{2}}a_n^{(1)}+a_n^{(2)})z^{-n}\rt),\\
&\Lambda_1(z)=:\exp\lt(\sum_{n\neq0}\frac{1}{n}\frac{1-t^{-n}}{1+p^{|n|}}z^n(a_{-n}^{(1)}-p^{\frac{|n|}{2}}a_{-n}^{(2)})\rt):,\\
&\Lambda_2(z)=:\exp\lt(-\sum_{n\neq0}\frac{1}{n}\frac{1-t^{-n}}{1+p^{|n|}}(\frac{z}{p})^n(a_{-n}^{(1)}-p^{\frac{|n|}{2}}a_{-n}^{(2)})\rt):=:\Lambda_1^{-1}\lt(\frac{z}{p}\rt):.
 \end{split}
 \end{equation}

\section{Relation with MO's R-matrix}\label{s:MO}

In this section, we first review the Maulik-Okounkov's R-amtrix \cite{MO}, which was induced from the instanton moduli space of 4d $\cN=2$ super Yang-Mills gauge theories, with a short-cut approach taken in \cite{ZM}, and then derive the key relation ${\bf R}^{MO}(u)T(Q,u)=T(-Q,u){\bf R}^{MO}(u)$ satisfied by MO's R-matrix from the horizontal representation of the universal R-matrix of DIM by taking the limit $q\rightarrow 1$. 

\subsection{A Brief Review on Maulik-Okounkov's (MO's) R-matrix}\label{review-MO}

MO's R-matrix is defined on the tensor product of two Fock spaces with background charge $\eta_1$ and $\eta_2$, i.e. the vector spaces are generated by bosons $\alpha^{(1)}_n$ and $\alpha^{(2)}_n$, satisfying $[\alpha^{(i)}_n,\alpha^{(j)}_m]=\delta_{ij}n\delta_{n+m,0}$, with the vacuum state of each Fock space being the eigenstate of the boson zero mode, $\alpha^{(i)}_0\ket{\eta_i}=\eta_i\ket{\eta_i}$, $\alpha_n^{(i)}\ket{\eta_i}=0$ for $n>0$ (no summation). 

The R-matrix only acts on the subspace spanned by $\alpha^-_n=\frac{1}{\sqrt{2}}\lt(\alpha^{(1)}_n-\alpha^{(2)}_n\rt)$ and the defining relation it satisfies is 
\ba
{\bf R}^{MO}(\bar{u})L_n(\bar{u},Q)=L_n(\bar{u},-Q){\bf R}^{MO}(\bar{u})\label{def-R-MO},
\ea
where $\bar{u}=\frac{1}{\sqrt{2}}(\eta_2-\eta_1)$ and 
\ba
L_n(\bar{u},Q)=\frac{1}{2}\sum_m{}':\alpha^-_{n+m}\alpha^-_{-m}:+Qn\alpha^-_n-\bar{u}\alpha^-_n=:L^{(0)}_n+Qn\alpha^-_n-\bar{u}\alpha^-_n.
\ea
$\sum{}'$ denotes the summation without terms involving the boson zero mode. The overall scale of the R-matrix is fixed by its action on the vacuum state, 
\ba
{\bf R}^{MO}(\bar{u})\ket{\eta_1}\otimes\ket{\eta_2}=\ket{\eta_1}\otimes\ket{\eta_2}.
\ea
In \cite{ZM}, it was argued that (\ref{def-R-MO}) solves the R-matrix uniquely with the above normalization. 

Once we obtained the explicit expression of MO's R-matrix, we can construct a family of lattice integrable systems by just using the R-matrix as the Lax operator. The transfer matrix in the $N$-site lattice model is 
\ba
T(\bar{u})={}_0\bra{\eta}{\bf R}_{01}^{MO}(\bar{u}){\bf R}_{02}^{MO}(\bar{u})\dots{\bf R}_{0N}^{MO}(\bar{u})\ket{\eta}_0,
\ea
where we take $\bar{u}=\frac{1}{\sqrt{2}}(\eta_i-\eta)$ for $i=1,\dots,N$. For $N=1$, we find the Hamiltonian of the system matching with the Calogero-Sutherland Hamiltonian with infinite number of particles \cite{AMOS,MimachiYamada}, 
\ba
{\cal H}=\sum_{n,m>0}(\alpha_{-n}\alpha_{-m}\alpha_{n+m}+\alpha_{-n-m}\alpha_n\alpha_m)+\sqrt{2}Q\sum_{n>0}n\alpha_{-n}\alpha_n,
\ea
where $\alpha_n$ is the boson on the first site. For larger $N$, we obtain a coproduct structure for the Calogero-Sutherland Hamiltonian. 

The Yangian algebra constructed from MO's R-matrix is isomorphic to the Yangian of $\widehat{\mathfrak{gl}_1}$ \cite{MO} (see also \cite{ZM} for a short review) and this Yangian algebra can also be obtained in the degenerate limit $q\rightarrow 1$ with $t=q^\beta$ for fixed $\beta$ from the DIM algebra \cite{Tsymbaliuk14}. 

\subsection{From R-matrix in DIM to MO's R-matrix}

In this section, we start from the defining condition of the universal R-matrix in DIM, 
\ba
{\bf R}\Delta(e)(z){\bf R}^{-1}=\cP\circ\Delta(e)(z),\ \ {\rm for}\ e\in U_{q_1,q_2}(\widehat{\mathfrak{gl}_1}),\label{Def-R}
\ea
to explain the origin of the defining conditions for MO's R-matrix by taking the horizontal representation of the universal R-matrix. The appearances of them are very different and it is not so obvious if they should be the same.  We show, however, that they agree with each other in the degenerate limit, where we set $q=\mathrm{e}^\hbar$, $t=q^{\beta}={\rm e}^{\beta\hbar}$, and take $\hbar\rightarrow 0$ limit. 

To compare with quantities in MO's context, we identify the weight $u$ in the horizontal representation as the $q$-boson zero mode $t^{a_0}$. Then using the notation $a^\pm_0=\frac{1}{\sqrt{2}}\lt(a_0^{(1)}\pm a_0^{(2)}\rt)$, we can rewrite the $q$-Virasoro operator as 
\begin{equation}
\begin{split}
 \rho_{u_1,u_2}(\Delta(t(z)))&=t^{a_0^{(1)}}\Lambda_1(z)+t^{a_0^{(2)}} \Lambda_2(z)\\&=t^{\frac{1}{\sqrt{2}}a_0^+}(t^{\frac{a_0^{-}}{\sqrt{2}}}\Lambda_1(z)+t^{-\frac{a_0^{-}}{\sqrt{2}}}\Lambda_2(z)).
\end{split}
 \end{equation}
By putting $t^{\frac{a_0^{-}}{\sqrt{2}}} \Lambda_1(z)=1+A(z)\hbar+B(z)\hbar^2+\cO(\hbar^3)$, and using $\Lambda_2(z)=:\Lambda_1^{-1}(\frac{z}{p}):$, we have
\ba
t^{-\frac{a_0^{-}}{\sqrt{2}}} \Lambda_2(z)=1-A\lt(\frac{z}{p}\rt)\hbar+\lt(:A^2\lt(\frac{z}{p}\rt):-B\lt(\frac{z}{p}\rt)\rt)\hbar^2+\cO(\hbar^3)\nn\\=1-A(z)\hbar+(:A^2(z):-B(z)-C(z))\hbar^2+\cO(\hbar^3),\nn
\ea
where, $A\lt(\frac{z}{p}\rt)=:A(z)+\hbar C(z)+\cO(\hbar^2)$.
Summarizing the above,  we get the following simple result:
\begin{equation}
 \rho_{u_1,u_2}(\Delta(t(z)))=t^{\frac{1}{\sqrt{2}}a_0^+}(2+(:A^2(z):-C(z))\hbar^2+\cO(\hbar^3)).
\end{equation} 
Expanding $\Lambda_1(z) $ with respect to $\hbar$, we see that 
 \begin{equation}
 A(z)=\frac{\beta}{\sqrt{2}}\sum_{n=-\infty}^{\infty}a_{-n}^{-}z^n,\quad
 C(z)=-\frac{\beta(1-\beta)}{\sqrt{2}}\sum_{n=-\infty}^{\infty}na_{-n}^{-}z^n,
\end{equation} 
where we used a similar notation $a^\pm_n=\frac{1}{\sqrt{2}}\lt(a^{(1)}_n\pm a^{(2)}_n\rt)$. 
Combined together, we have 
\begin{equation}
 :A^2(z):-C(z)=\beta\left(\frac{\beta}{2}\sum_{m,n=-\infty}^{\infty}:a_{-n}^{-}a_{-m}^{-}:z^{n+m}+\frac{1-\beta}{\sqrt{2}}\sum_{n=-\infty}^{\infty}na_{-n}^{-}z^n\right).
\end{equation}
In the $\hbar\rightarrow 0$ limit, $a_n$ can be identified with $\frac{1}{\sqrt{\beta}}\alpha_n+\cO(\hbar)$, and at the leading order the $q$-Virasoro operator becomes, 
\ba
\rho_{u_1,u_2}(\Delta(t(z))) =t^{\frac{1}{\sqrt{2\beta}}\alpha_0^+}\left(2+\hbar^2\beta\sum_{n=-\infty}^{\infty}(L^{(0)}_{-n}+Qn\alpha^{-}_{-n}+\alpha_0^{-}\alpha_{-n}^{-})z^n+\cO(\hbar^3)\right)\nn\\
=:t^{\frac{1}{\sqrt{2\beta}}\alpha_0^+}(2+\hbar^2\beta T(Q,\bar{u})+\cO(\hbar^3)),
\ea
where, $Q=-\frac{1}{\sqrt{2}}(\sqrt{\beta}-\frac{1}{\sqrt{\beta}})$, $L^{(0)}_{-n}=\frac{1}{2}\sum_m{}':\alpha_{n-m}^{-}\alpha_m^{-}:$. 

Since $\cP$ exchanges the d.o.f.'s on two sites, it flips the sign of $\alpha^-_n$, i.e. $\cP\alpha^-_n=-\alpha^-_n$ and thus turns $T(Q,\bar{u})$ into $T(-Q,\bar{u})$\footnote{Rigorously speaking, we should write it as $\rho_{u_1,u_2}({\bf R}(u))\rho_{u_1,u_2}(\Delta(t(z)))=\lt(\cP\cdot\rho_{u_1,u_2}(\Delta(t(z)))\rt)\rho_{u_1,u_2}({\bf R}(u))$, but to avoid the complexity of the equations, we omit the representation map for ${\bf R}(u)$. It should be understood that all ${\bf R}(u)$'s appearing in this article is in the horizontal representation.}. 
\ba
&&{\bf R}(u)\rho_{u_1,u_2}(\Delta(t(z)))=\lt(\cP\cdot\rho_{u_1,u_2}(\Delta(t(z)))\rt){\bf R}(u)\Rightarrow\nn\\
&&{\bf R}(u)\lt(2+\hbar^2\beta T(Q,\bar{u})+\cO(\hbar^3)\rt)=\lt(2+\hbar^2\beta T(-Q,\bar{u})+\cO(\hbar^3)\rt){\bf R}(u),
\ea
where the natural choice of the spectral parameter is $u=\frac{u_1}{u_2}=t^{-\frac{1}{\sqrt{2\beta}}\bar{u}}$. We expand ${\bf R}(u)=\sum_{n\geq0}R^{(n)}(\bar{u})\hbar^n$, then we see that the leading order of the R-matrix satisfies 
\ba
R^{(0)}T(Q,\bar{u}) &= T(-Q,\bar{u}) R^{(0)},
\ea
and it can be identified as MO's R-matrix ${\bf R}^{MO}(\bar{u})$. 
 
\paragraph{Remark}

The relation ${\bf R}\Delta(e)=\cP\circ\Delta(e){\bf R}$ is satisfied by all the DIM elements $e$. The DIM algebra is generated by $x^{\pm}(z)$, $b_n$ and centers. 
The centers are not relevant in the R-matrix relations. The other generators of the DIM algebra are $t(z)$, $t^\ast(z)$ and $b_n$, where 
\begin{equation}
 \begin{split}
  & t^*(z)=\alpha(p^{-1}z)^{-1}x^-(p^{-1}\gamma^{-1}z)\beta(\gamma^{-2}p^{-1}z)^{-1},\\
   &\rho(\Delta(t^*(z)))=t^{a_0^{(2)}}\Lambda_2(z)+t^{a_0^{(1)}}\Lambda_1(z)=\rho(\Delta(t(z))).
 \end{split}
\end{equation}
We see that $t^\ast(z)$ induces the same defining relation for the R-matrix. 
The remaining relation we have not considered is ${\bf R}\Delta(b_n) = \cP\circ\Delta(b_n){\bf R}$, which in the degenerate limit $\hbar\to 0$, reduces to $R^{(0)}\alpha_n^+=\alpha_n^+R^{(0)}$. This explains why MO's R-matrix only acts on the diagonal space spanned by $\alpha^-$.

 \section{Explicit Computation of R-matrix}\label{Exp-R}
In this section, we solve the defining equation of the universal R-matrix in the horizontal representation in terms of the $q$-boson oscillators at the first two lowest orders. For simplicity, we omit the notation for the horizontal representation map $\rho$  and we use the short-hand notation $\Delta^{op}=\cP\circ\Delta$ in the following. 

Unlike in the degenerate case of MO's R-matrix, the constraint ${\bf R}\Delta(b_n)=\Delta^{op}(b_n){\bf R}$, namely ${\bf R}(p^{\frac{|n|}{2}}a_n^{(1)}+a_n^{(2)})=(p^{\frac{|n|}{2}}a_n^{(2)}+a_n^{(1)}){\bf R}$ is less trivial. This equation does not depend on the spectral parameter, so we have the folllowing operator, which exchanges $a_n^{(1)}$ and $a_n^{(2)}$, as part of the R-matrix,  
\begin{equation}
R_{--}=\exp\lt(\pi i\sum_{m=1}^{\infty}\frac{1}{m}\frac{1-t^m}{1-q^m}a_{-m}^{-}a_m^{-}\rt).
\end{equation} 
With the Campbell-Baker-Hausdorff formula (see Appendix \ref{A} for a review and derivation), we have 
\begin{equation}
   R_{--}a_n^{(1)}R_{--}^{-1}=a_n^{(2)}, \quad R_{--}a_n^{(2)}R_{--}^{-1}=a_n^{(1)} \quad (\mathrm{for}\; n\neq0).\label{check-CBH}
\end{equation}  
 
 Let us put an ansatz for the R-matrix ${\bf R}(u)={\bf R}'(u)R_{--}$, then  we have ${\bf R}'(u)\lt(p^{\frac{|n|}{2}}a_n^{(2)}+a_n^{(1)}\rt)=\lt(p^{\frac{|n|}{2}}a_n^{(2)}+a_n^{(1)}\rt){\bf R}'(u)$. This means ${\bf R}'$ commutes with $p^{\frac{|n|}{2}}a_n^{(2)}+a_n^{(1)}$, so ${\bf R}'$ is constructed only from the boson operator $a^{p-}_n:=a_n^{(2)}-p^{\frac{|n|}{2}}a_n^{(1)}$.
  
The defining relation now becomes 
\begin{equation}
 \begin{split}
{\bf R}'\Delta^{op'}(t(z)){\bf R}'^{-1}=\Delta^{op}(t(z)),
 \end{split}
\end{equation}
where, $\Delta^{op'}(t(z))$ and $\Delta^{op}(t(z))$ reads ($^{op'}$ means that we do not exchange $u_1$ and $u_2$ as in $\Delta^{op}$)
 \begin{equation}
  \begin{split}
   & \Delta^{op'}(t(z)) =u_1\Lambda_1^{op}(z)+u_2 \Lambda_2^{op}(z),\\
  & \Delta^{op}(t(z))) =u_2\Lambda_1^{op}(z)+u_1\Lambda_2^{op}(z).
 \end{split}
 \end{equation}
Now $\Lambda_1^{op}(z), \Lambda_2^{op}(z)$ after the action of $\cP$ read 
\begin{equation}
\begin{split}
&\Lambda_1^{op}(z)=:\exp\lt(\sum_{n\neq0}\frac{1}{n}\frac{1-t^{-n}}{1+p^{|n|}}z^n(a_{-n}^{(2)}-p^{\frac{|n|}{2}}a_{-n}^{(1)})\rt):,\\
&\Lambda_2^{op}(z)=:\exp\lt(-\sum_{n\neq0}\frac{1}{n}\frac{1-t^{-n}}{1+p^{|n|}}\lt(\frac{z}{p}\rt)^n(a_{-n}^{(2)}-p^{\frac{|n|}{2}}a_{-n}^{(1)})\rt):=\Lambda_1^{op}\lt(\frac{z}{p}\rt)^{-1}.
 \end{split}
 \end{equation} 
Recall that $u=\frac{u_1}{u_2}$, we have  
\begin{equation} 
 {\bf R}'(u)\left(u\Lambda_1^{op}(z)+\Lambda_2^{op}(z)\right){\bf R}'^{-1}(u)=\Lambda_1^{op}(z)+u\Lambda_2^{op}(z).
 \end{equation}
At $u=0$, the above equation becomes ${\bf R}'(0)\Lambda_2^{op}(z){\bf R}'(0)^{-1}=\Lambda_1^{op}(z)$ and the solution is
\begin{equation}
   {\bf R}'(0)=\mathrm{exp}\left(\sum_{m=1}^{\infty}\lt(\hbar(1-\beta)+\frac{\pi i}{m}\rt)\frac{1-t^m}{(1-q^m)(1+p^m)}a^{p-}_{-m}a^{p-}_m\right).
\end{equation} 
This follows from the fact ${\bf R}'(0)a^{p-}_{-n}{\bf R}'^{-1}(0)=-p^na^{p-}_{-n}$. It also satisfies the property 
\begin{equation}
  {\bf R}'(0)\left(u\Lambda_1^{op}(z)+\Lambda_2^{op}(z)\right){\bf R}'^{-1}(0)=\Lambda_1^{op}(z)+u\Lambda_2^{op}(p^2z).
\end{equation} 
Since ${\bf R}'(u)$ is fixed for small $u$ now, one may expand it in terms of $u$ around $u=0$.  We write ${\bf R}'(u)$ as 
\ba
{\bf R}'(u)=\lt(1+R^{(1)}u+R^{(2)}u^2+\cdots\rt){\bf R}'(0),
\ea
then the equation we need to solve becomes
\begin{equation} 
 (1+R^{(1)}u+R^{(2)}u^2+\cdots)\left(\Lambda_1^{op}(z)+u\Lambda_2^{op}(p^2z)\right)=(\Lambda_1^{op}(z)+u\Lambda_2^{op}(z)) (1+R^{(1)}u+R^{(2)}u^2+\cdots),\nn
\end{equation}
which reduces to the recursive equation, 
\begin{equation}
  \lt[R^{(n)}, \Lambda_1^{op}(z)\rt]=\Lambda_2^{op}(z)R^{(n-1)}-R^{(n-1)}\Lambda_2^{op}(p^2z).
\end{equation}
For $n=1$, $\lt[R^{(1)}, \Lambda_1^{op}(z)\rt]=\Lambda_2^{op}(z)-\Lambda_2^{op}(p^2z)$. Our claim is that the solution is given by 
\begin{equation}
\begin{split}
  uR^{(1)}&=\frac{(1-p)} {(1-q)(1-t^{-1})}\lt(\sum_{r\in\mathbb{Z}}x^+_r\otimes x^-_{-r}-u\rt)\\
&=\frac{u(1-p)}{(1-q)(1-t^{-1})}\lt(\lt. :\exp\lt(-\sum_{n\neq0}\frac{1-t^{-n}}{n}p^{-\frac{|n|}{2}}a^{p-}_{-n}w^n\rt):\rt|_{\mathrm{zero\ mode}}-1\rt),
\end{split}
\end{equation}
where the normalization of the R-matrix is again fixed by ${\bf R}(u)\ket{0}\otimes\ket{0}=\ket{0}\otimes\ket{0}$, as we required in the case of MO's R-matrix. 

\begin{proof}
\ \\
 In general, if $[A, B] $ is c-number, the equation $\mathrm{e}^A\mathrm{e}^B=\mathrm{e}^{[A, B]}\mathrm{e}^{B}\mathrm{e}^A$ holds. Using this, we can show the equation $\lt[:\mathrm{e}^{A(w)}:,:\mathrm{e}^{B(z)}:\rt]=\lt(\mathrm{e}^{[A_+(w),\  B_-(z)]}-\mathrm{e}^{[B_+(z),\  A_-(w)]}\rt):\mathrm{e}^{A(w)+B(z)}:$, where $A(w)$ and $B(z)$ are linear summation of boson oscillators and $\pm$ denotes the positive/negative mode part.
  
With this identity, we have 
\ba
   &&\lt[:\exp\lt(-\sum_{n\neq0}\frac{1-t^{-n}}{n}p^{-\frac{|n|}{2}}a^{p-}_{-n}w^n\rt): , \Lambda_1^{op}(z)\rt]\nn\\
   &&=\left(\exp\left(\sum_{n=1}^{\infty}\frac{(1-q^n)(1-t^{-n})}{np^{\frac{n}{2}}}\lt(\frac{z}{w}\rt)^n\right)-\exp\left(\sum_{n=1}^{\infty}\frac{(1-q^n)(1-t^{-n})}{np^{\frac{n}{2}}}\lt(\frac{w}{z}\rt)^n\right)\right)\nn\\
 &&\hspace{160pt}:\mathrm{exp}\left(\sum_{n\neq0}\frac{1}{n}\frac{1-t^{-n}}{1+p^{|n|}}\biggl(z^n-(p^{\frac{|n|}{2}}+p^{-\frac{|n|}{2}})w^n\biggr)a^{p-}_{-n}\right):.
\ea
  
Note that 
\ba
  &&\mathrm{exp}\left(\sum_{n=1}^{\infty}\frac{(1-q^n)(1-t^{-n})}{np^{\frac{n}{2}}}x^n\right)
  =\frac{(1-qp^{-\frac{1}{2}}x)(1-t^{-1}p^{-\frac{1}{2}}x)}{(1-p^{-\frac{1}{2}}x)(1-p^{\frac{1}{2}}x)}\nn\\
  &&=1+\frac{(1-q)(1-t^{-1})}{(1-p)}\biggl(\frac{1}{1-p^{-\frac{1}{2}}x}-\frac{1}{1-p^{\frac{1}{2}}x}\biggr),\nn
\ea
(it only applies around $x\sim 0$) we have
\begin{equation}
  \begin{split}
   &\quad\frac{(1-p)}{(1-q)(1-t^{-1})}\lt[:\mathrm{exp}(-\sum_{n\neq0}\frac{1-t^{-n}}{n}p^{-\frac{|n|}{2}}a^{p-}_{-n}w^n): , \Lambda_1^{op}(z)\rt]\\
   =&\biggl(\frac{1}{1-p^{-\frac{1}{2}}\frac{z}{w}}-\frac{1}{1-p^{\frac{1}{2}}\frac{z}{w}}-\frac{1}{1-(p^{\frac{1}{2}}\frac{z}{w})^{-1}}+\frac{1}{1-(p^{-\frac{1}{2}}\frac{z}{w})^{-1}}\biggr):\mathrm{exp}\left(\sum_{n\neq0}\frac{1}{n}\frac{1-t^{-n}}{1+p^{|n|}}\biggl(z^n-(p^{\frac{n}{2}}+p^{-\frac{n}{2}})w^n\biggr)a^{p-}_{-n}\right):\\
=&\left(\delta\lt(\frac{z}{p^{\frac{1}{2}}w}\rt)-\delta\lt(\frac{p^{\frac{1}{2}}z}{w}\rt)\right):\mathrm{exp}\left(\sum_{n\neq0}\frac{1}{n}\frac{1-t^{-n}}{1+p^{|n|}}\biggl(z^n-(p^{\frac{n}{2}}+p^{-\frac{n}{2}})w^n\biggr)a^{p-}_{-n}\right):\\
=&\delta\lt(\frac{z}{p^{\frac{1}{2}}w}\rt):\mathrm{exp}\left(-\sum_{n\neq0}\frac{1}{n}\frac{1-t^{-n}}{1+p^{|n|}}\biggl(\frac{z}{p}\biggr)^na^{p-}_{-n}\right):-\delta\lt(\frac{p^{\frac{1}{2}}z}{w}\rt):\mathrm{exp}\left(-\sum_{n\neq0}\frac{1}{n}\frac{1-t^{-n}}{1+p^{|n|}}\bigl(zp\bigr)^na^{p-}_{-n}\right):\\
=&\delta\lt(\frac{z}{p^{\frac{1}{2}}w}\rt)\Lambda_2^{op}(z)-\delta\lt(\frac{p^{\frac{1}{2}}z}{w}\rt)\Lambda_2^{op}(p^2z).\nn
 \end{split}
 \end{equation}  

The zero-mode ($w^0$-order term) is exactly $ \Lambda_2^{op}(z)-\Lambda_2^{op}(p^2z)$. 

\end{proof}

Higher orders can be determined from the recursive formula, but it is rather complicated to perform the calculation. 

\section{Charges of the Associated Integrable System}\label{s:charges}

Let us compute the first charge from $R^{(1)}$ imitating the calculation in the degenerate limit reviewed in section \ref{review-MO}. The monodoromy operator is ${\bf T}_0(u)={\bf R}_{10}(u){\bf R}_{20}(u)\dots{\bf R}_{N0}(u)$. The $N$-site transfer matrix is given by 
\begin{equation}
t_N(u):={}_0\bra{0}{\bf T}(u)\ket{0}_0={}_0\bra{0}{\bf R}_{10}(u){\bf R}_{20}(u)\dots {\bf R}_{N0}(u)\ket{0}_0.
\end{equation}

In the one-site case, the transfer matrix reads 
\begin{equation}
 {}_0\bra{0}{\bf R}_{10}(u)\ket{0}_0={}_0\bra{0}{\bf R}'(0)R_{--}\ket{0}_0+u\ {}_0\bra{0}R^{(1)}{\bf R}'(0)R_{--}\ket{0}_0+\dots\nn
\end{equation}
The first term gives rise to some vertex operator, which is a overall factor of the R-matrix. The second term can be obtained from the $w^0$-order term of 
\ba
&&{x^{+(1)}}(w)\ {}_0\bra{0}{x^{-(0)}}(w){\bf R}'(0)R_{--}\ket{0}_0\nn \\
&&= {x^{+(1)}}(w)\ {}_0\bra{0}:\exp\lt(-\sum_{n\neq0}\frac{1-t^{-n}}{n}p^{-\frac{|n|}{2}}a_{-n}^{(0)}z^n\rt):{\bf R}'(0)R_{--}\ket{0}_0\nn\\
&&={x^{+(1)}}(w)\ {}_0\bra{0}{\bf R}'(0)R_{--}\ket{0}_0,
\ea
where we used 
\begin{equation}
\begin{split}
a_n^{(0)}{\bf R}_{10}'(0)R_{--10}&=\frac{1}{p^{\frac{n}{2}}+p^{-\frac{n}{2}}}{\bf R}'_{10}(0)a_n^{(1)}R_{--10}\\
&=\frac{1}{p^{\frac{n}{2}}+p^{-\frac{n}{2}}}{\bf R}'_{10}(0)R_{--10}\ a_n^{(0)}\ \ \ \ (\mathrm{for}\  n\geq1).\nn
\end{split}
\end{equation}
Dividing out the overall factor, we obtain the first chrage $x^+_0$, which agrees with the Hamiltonian of  the Ruijsenaars-Schneider model \cite{boson-RS}.

It is easy to find charges in the case of $N$-site. Let us decompose the R-matrix as ${\bf R}(u)=\sum_ia_i\otimes b_i$, then for example, charges in the case of 2-site can be computed as follows. 
\begin{equation}
\begin{split}
&{}_0\bra{0}{\bf R}_{10}(u){\bf R}_{20}(u)\ket{0}_0\\
&={}_0\bra{0}(\Delta\otimes 1){\bf R}_{12}(u)\ket{0}_0\\
&=\sum_i\Delta(a_i)\ {}_0\bra{0}b_i^{(0)}\ket{0}_0\\
&=\Delta\left({}_0\bra{0}R_{10}(u)\ket{0}_0\right),\nn
\end{split}
\end{equation}
where in the last line we used the fact that ${}_0\bra{0}b_i^{(0)}\ket{0}_0$'s are merely $c$-numbers. This result suggests that charges in the two-site case are coproduct of charges in the case of one-site. The two-site Hamiltonian\footnote{We note that this is a conjecture. Even though the first charge is computed to have the same expression as $x^+_0$, we did not prove that it has the same coproduct as $x^+_0$ as we have already took the horizontal representation and expressed everything in the $q$-boson, which has no definite coproduct structure.} is $\Delta(x^+_0)=\left.x^+(z)\otimes1+\psi^-(\gamma_{(1)}^{\frac{1}{2}}z)\otimes x^+(\gamma_{(1)}z)\right|_{\rm zero mode}$ and the coproduct gives rise to non-trivial interactions between two sites. This argument can be generalized to higher $N$-site cases.

$x^+_0$ is exactly the S-dual of the first non-trivial charge in the vertical representation (or (0,$m$) representation, for $m\in\mathbb{Z}_{>0}$) \cite{DIMBethe}, therefore it is natural to conjecture that higher-rank charges correspond to the S-dual of $\psi^+_n$'s ($n\geq2$) in the vertical representation. 

\section{Comparison with the Results from \cite{MNagoya,D-Ohkubo}}\label{Ohkubo}

In this section, we compute the matrix elements of the oscillator expression obtained for ${\bf R}(u)$ to compare it with the known results in the literature \cite{MNagoya,D-Ohkubo}. 

Let $\ket{\lambda_1,\lambda_2}$ be the eigenstate\footnote{This basis is often called the generalized Macdonald polynomial basis. The normalization is determined by $\ket{\lambda_1,\lambda_2}=a_{-\lambda_1}^{(1)}a_{-\lambda_2}^{(2)}\ket{0}_1\otimes\ket{0}_2+\cdots$.}
 of $\Delta(x^+_0)$, i.e. $\Delta(x^+_0)\ket{\lambda_1,\lambda_2}=\kappa_{\lambda_1\lambda_2}\ket{\lambda_1,\lambda_2}$ with the corresponding eigenvalue $\kappa_{\lambda_1\lambda_2}$, then ${\bf R}(u)\ket{\lambda_1,\lambda_2}$ will be the eigenstate to $\Delta^{op}(x^+_0)$. 
\ba
\kappa_{\lambda_1\lambda_2}{\bf R}(u)\ket{\lambda_1,\lambda_2}={\bf R}(u)\Delta(x^+_0)\ket{\lambda_1,\lambda_2}=\Delta^{op}(x^+_0){\bf R}(u)\ket{\lambda_1,\lambda_2},
\ea
following from the definition of the universal R-matrix. From this relation, we have ${\bf R}(u)\ket{\lambda_1,\lambda_2}=R_{\lambda_1 \lambda_2}(u)\ket{\lambda_1,\lambda_2}^{\mathrm{op}}$. Here, $R_{\lambda_1 \lambda_2}(u)$ is a constant factor and $\ket{\lambda_1,\lambda_2}^{\mathrm{op}}$ is the polynomial obtained from $\ket{\lambda_1,\lambda_2}$ by replacing $u$ with $1/u$, and exchanging $a^{(1)}$ and $a^{(2)}$. 
In \cite{MNagoya}, the form of $R_{\lambda_1 \lambda_2}(u)$ is conjectured as the following;
\begin{equation}
\begin{split}
&R_{\lambda_1\lambda_2}(u)=(\frac{q}{t})^{\frac{1}{2}(|\lambda_1|+|\lambda_2|)}\frac{G_{\lambda_1 \lambda_2(u)}}{G_{\lambda_1 \lambda_2(\frac{q}{t}u)}}, \\
&G_{\lambda_1 \lambda_2(u)}=\Pi_{(i,j)\in\lambda_1}(1-uq^{\lambda_i^{(1)}-j}t^{{\lambda_j^{(2)}}^T-i+1}) \Pi_{(i,j)\in\lambda_2}(1-uq^{-\lambda_i^{(2)}+j-1}t^{{-\lambda_j^{(1)}}^T+i}).
\end{split}
\end{equation}
It was checked at low levels and was further studied in \cite{MNagoya2}.

Let us check the consistency with our result.
We apply the bosonic representation of the R-matrix obtained in section \ref{Exp-R} to $\ket{\lambda_1,\lambda_2}$ in the horizontal repsentation up to level $2$. 

At level $1$, 
there are two generalized Macdonald polynomials as following.\footnote{The label of the generalized Macdonald polynomial is determined so that the eigenvalue takes the form 
\ba
\kappa_{\lambda_1\lambda_2}=u_1e_{\lambda_1}+u_2e_{\lambda_2},\nn
\ea
with $e_\lambda=1+(t-1)\sum_i(q^{\lambda^{(i)}}-1)t^{-i}$ for $\lambda=\{\lambda^{(i)}\}$. 
This formula is quoted from \cite{D-Ohkubo}.}
\ba
&&\ket{\tyng(1),\emptyset}=a^{(1)}_{-1}\ket{0}_1\otimes\ket{0}_2,\nn\\
&&\ket{\emptyset,\tyng(1)}=-\frac{1-p}{\sqrt{p}(1-u)}a^{(1)}_{-1}\ket{0}_1\otimes\ket{0}_2+a^{(2)}_{-1}\ket{0}_1\otimes\ket{0}_2.\nn
\ea
The corresponding eigenvalues are 
\ba
\kappa_{\tyng(1)\emptyset}=u_2\lt(1+(t^{-1}-p+q)u\rt),\quad \kappa_{\emptyset\tyng(1)}=u_1\lt(1+(t^{-1}-p+q)u^{-1}\rt).
\ea
The action of ${\bf R}(u)$ is evaluated as 
\ba
\lt(1+R^{(1)}u+\cO(u^2)\rt){\bf R}'(0)R_{--}\ket{\tyng(1),\emptyset}=p^{1/2}\lt(1-(1-p)u\rt)a^{(1)}_{-1}\ket{0}_1\otimes\ket{0}_2+(1-p)ua^{(2)}_{-1}\ket{0}_1\otimes\ket{0}_2+\cO(u^2)\nn\\
=\frac{p^{-1/2}(u-1)}{u-1/p}\ket{\emptyset,\tyng(1)}^{op}+\cO(u^2),\nn\\
\lt(1+R^{(1)}u+\cO(u^2)\rt){\bf R}'(0)R_{--}\ket{\emptyset,\tyng(1)}=\lt((1+u)p^{1/2}-p^{-1/2}\rt)a^{(2)}_{-1}\ket{0}_1\otimes\ket{0}_2+\cO(u^2)\nn\\
=\frac{p^{-1/2}(u-p)}{u-1}\ket{\tyng(1),\emptyset}^{op}+\cO(u^2),\nn
\ea
The coefficients before $\ket{\lambda_1,\lambda_2}^{op}$ obtained here agree with the results  in \cite{MNagoya,D-Ohkubo}
 \  up to $\cO(u^2)$ .
One may also check such consistency with \cite{MNagoya,D-Ohkubo} at level 2. The details are explained in Appendix B.

\section{Conclusion and Discussion}

We showed in this paper that by taking the $(1,0)$ horizontal representation of the DIM algebra, MO's R-matrix can be obtained from the bosonic realization of the universal R-matrix in the degenerate limit $q\rightarrow 1$. The universal R-matrix, by construction, is associated with the algebra, not its representation, and in section \ref{Ohkubo}, we did see that the matrix elements and eigenvalues of the R-matrix computed in two ways agree with each other. We may also have MO's R-matrix as the degenerate limit of the universal R-matrix in a different representation, for example in the $(0,1)$ vertical representation. The degenerate limit of the DIM algebra, i.e. the $\widehat{\mathfrak{gl}_1}$ Yangian or the SH$^c$ algebra, however, does not present a manifest SL(2,$\mathbb{Z}$) symmetry, and it is unclear what is the suitable definition for MO's R-matrix in such a representation. We will left this problem to a future work. 

There are also a lot of questions unsolved even in the $(1,0)$ representation, due to the complexity of the calculation. We list them here as future directions to conclude this article. 
\begin{itemize}
\item We only solved one defining equation among three for the universal R-matrix, i.e. 
\ba
(\Delta\otimes 1){\bf R}={\bf R}_{13}{\bf R}_{23},\quad (1\otimes \Delta){\bf R}={\bf R}_{13}{\bf R}_{12},\label{def-R-A}
\ea
are left untouched. In fact, if we decompose the action of the R-matrix on the vacuum state as 
\ba
{\bf R}(u)\ket{0}_1\otimes \ket{0}_2=\sum_i a_i\otimes b_i\ket{0}_1\otimes \ket{0}_2,\nn
\ea
using (\ref{def-R-A}), we can show that ${\bf R}(u)$ can only act trivially on the vacuum, i.e. ${\bf R}(u)\ket{0}_1\otimes \ket{0}_2=\ket{0}_1\otimes \ket{0}_2$. We conjecture that by setting such a normalization condition, the solution of R-matrix obtained in this article satisfies (\ref{def-R-A}), as there is only one unique solution to (\ref{Def-R}) with this specific normalization. We emphasize again that it is not straightforward to check this conjecture even at the leading order in $u$ of the R-matrix. This is due to the fact that we took the horizontal representation of the R-matrix and we cannot assign a coproduct for the $q$-boson used in the representation. 

\item It is not clear how the oscillator expression of the universal R-matrix is related to that of MO's R-matrix. The recursive formula obtained in this paper for the universal R-matrix is not directly related to that for MO's R-matrix. As we can see that $R^{(1)}$ starts at order $\cO(\hbar)$ in the degenerate limit, which also implicates that all $R^{(n)}$ ($n\geq 1$) are of order $\cO(\hbar)$, the naive limit $\hbar\rightarrow 0$ suggests 
\ba
\lim_{\hbar\rightarrow 0}{\bf R}(u)=1,\nn
\ea
which is clearly wrong. Interestingly, at the leading order of $\hbar$, the recursive equation for $R^{(n)}$'s becomes trivial ($\lt[R^{(n+1)},\alpha^-_{k}\rt]=\lt[R^{(n)},\alpha^-_{k}\rt]$), and the leading order in $\hbar$ of $R^{(1)}$, $(1-\beta)\hbar\sum_{n>0}\alpha_{-n}\alpha_n$, survives in all $R^{(n)}$'s. We can resum all of them into an $\cO(\hbar^0)$ term, 
\ba
\sum_{k=1}^\infty (1-\beta)\hbar u^k\sum_{n>0}\alpha_{-n}\alpha_n=\frac{(1-\beta)\hbar u}{1-u}\sum_{n>0}\alpha_{-n}\alpha_n\rightarrow \frac{2Q}{\bar{u}}\sum_{n>0}\alpha_{-n}\alpha_n,
\ea
which reproduces the first non-trivial term in the expansion of MO's R-matrix (for example see \cite{ZM}). In other words, this suggests that $R^{(1)}$ contains almost all pieces of higher rank terms of MO's R-matrix as its $\hbar$-expansion, with corrections from $R^{(n)}$'s ($n>1$). Unfortunately, we do not know how to reproduce the second non-trivial expansion mode in MO's R-matrix in a similar way at the current stage. 

\item We neglected an overall vertex operator ${\bf R}'(0)R_{--}$ in the calculation of charges in the related integrable lattice model. It is not clear what kind of role it plays both 
in  Ruijsenaars-Schneider model and in the degenerate limit. 

\end{itemize}

\section*{Acknowledgement}
We thank J.-E. Bourgine for helpful discussions and comments on the manuscript. YM is partially supported by Grants-in-Aid for Scientific Research (Kakenhi $\#$25400246) from MEXT, Japan. MF and RZ are supported by JSPS fellowship. 

\appendix

\section{Campbell-Baker-Hausdorff formula}\label{A}

Let us check equation (\ref{check-CBH}). We use the following Campbell-Baker-Hausdorff formula, where $(\mathrm{ad}A)B=[A,B]$. 
\begin{equation}
\mathrm{e}^AB\mathrm{e}^{-A}=B+[A,B]+\frac{1}{2}[A,[A,B]]+\cdots=\mathrm{e}^{\mathrm{ad}A}B.
\end{equation}
Using  relation $[a_m^{(-)}, a_n^{(-)}] = m\frac{1-q^{|m|}}{1-t^{|m|}}\delta_{m+n,0}$, we have the following equation.
\begin{equation}
\begin{split}
&[\pi i\sum_{m=1}^{\infty}\frac{1}{m}\frac{1-t^m}{1-q^m}a_{-m}^{(-)}a_m^{(-)}, a_n^{(-)}]=-\pi ia_n^{(-)},\\
&[\pi i\sum_{m=1}^{\infty}\frac{1}{m}\frac{1-t^m}{1-q^m}a_{-m}^{(-)}a_m^{(-)}, a_{-n}^{(-)}]=\pi ia_{-n}^{(-)}\hspace{20pt} (  \mathrm{for}\  n>0\ ).
\end{split}
\end{equation}

Using this equation and Campbell-Baker-Hausdorff formula, we have the following equation.
\begin{equation}
\begin{split}
&\hspace{30pt}R_{--}a_n^{(-)}R_{--}^{-1}\\
=\hspace{20pt}&\mathrm{exp}\lt(\pi i\sum_{m=1}^{\infty}\frac{1}{m}\frac{1-t^m}{1-q^m}a_{-m}^{(-)}a_m^{(-)}\rt)a_n^{(-)}\mathrm{exp}\lt(-\pi i\sum_{m=1}^{\infty}\frac{1}{m}\frac{1-t^m}{1-q^m}a_{-m}^{(-)}a_m^{(-)}\rt)\\=\hspace{20pt}&\hspace{40pt}\mathrm{e}^{\pm\pi i}a_n^{(-)}\\=\hspace{20pt}&\hspace{40pt}-a_n^{(-)}\hspace{20pt}(\ \mathrm{for}\ n\neq0).
\end{split}
\end{equation}

$R_{--}a_n^{(+)}R_{--}^{-1}=a_n^{(+)}$ holds because of the trivial commutation relation $[R_{--}, a_n^{(+)}]=0$.

Combining these relations between $R_{--}$ and $a_n^{(\pm)}$, we have equation (\ref{check-CBH}).

\section{Matrix Element of the R-matrix at Level Two}\label{B}

At level two, the generalized Macdonald basis is known as 
\ba
&&\ket{\tyng(1,1),\emptyset}=\frac{(1+q)(1-t)}{1-q)(1+t)}a^{(1)}_{-1}a^{(1)}_{-1}\ket{0}_1\otimes\ket{0}_2+a^{(1)}_{-2}\ket{0}_1\otimes\ket{0}_2,\nn\\
&&\ket{\tyng(2),\emptyset}=a^{(1)}_{-1}a^{(1)}_{-1}\ket{0}_1\otimes\ket{0}_2-a^{(1)}_{-2}\ket{0}_1\otimes\ket{0}_2,\nn\\
&&\ket{\tyng(1),\tyng(1)}=\frac{(1-p)(2t-(1+q-t+qt))}{2p^{1/2}(1-qu)(u-t)}a^{(1)}_{-1}a^{(1)}_{-1}\ket{0}_1\otimes\ket{0}_2+\frac{(1-p)(1-q)(1+t)u}{2p^{1/2}(qu-1)(u-t)}a^{(1)}_{-2}\ket{0}_1\otimes\ket{0}_2\nn\\
&&+a^{(1)}_{-1}a^{(2)}_{-1}\ket{0}_1\otimes\ket{0}_2,\nn
\ea
\ba
&&\ket{\emptyset,\tyng(2)}=\frac{(1-p)\lt((1-p)(1+t)-(1+q+t-2pt-pqt-t^2+pt^2)u+(t+qt-t^2-qt^2)u^2\rt)}{p(1-u)(1-tu)(1-(u+qu-1)t)}\nn\\
&&\times a^{(1)}_{-1}a^{(1)}_{-1}\ket{0}_1\otimes\ket{0}_2\nn\\
&&+\frac{(1-p)\lt((1+p)(1+t)-(1-q+3t+2pt+pqt+t^2+pt^2)u+(t-qt+t^2+2pt^2+qt^2)u^2\rt)}{p(1-u)(1-tu)(1-t(u+qu-1))}\nn\\
&&\times a^{(1)}_{-2}\ket{0}_1\otimes\ket{0}_2\nn\\
&&+\frac{2(1-p)}{p^{1/2}(1-tu)}a^{(1)}_{-1}a^{(2)}_{-1}\ket{0}_1\otimes\ket{0}_2+a^{(2)}_{-1}a^{(2)}_{-1}\ket{0}_1\otimes\ket{0}_2-a^{(2)}_{-2}\ket{0}_1\otimes\ket{0}_2,\nn
\ea
\ba
&&\ket{\emptyset,\tyng(1,1)}=\frac{(1-p)(1+q)(1-t)}{p(1-q)(1+t)(1-u)(q-u)((1+t)u-(1+q)t)}\lt((1-q)(1+t)u^2\rt.\nn\\
&&\lt.-\lt(t-qt-q^2(1+t)+p(t-q(1+2t))\rt)u-qt(1-p)(1+q)\rt)a^{(1)}_{-1}a^{(1)}_{-1}\ket{0}_1\otimes\ket{0}_2\nn\\
&&-\frac{(1-p)}{p(1-u)(u-q)((1+t)u-(1+q)t)}\lt((1-q(1-t)+t+2pt)u^2\rt.\nn\\
&&\lt.+((1-t)q^2-t-3qt-pq-pt-2pqt)u+qt(1+p)(1+q)\rt)a^{(1)}_{-2}\ket{0}_1\otimes\ket{0}_2\nn\\
&&+\frac{2q(1-p)(1+q)(1-t)}{p^{1/2}(1-q)(1+t)(u-q)}a^{(1)}_{-1}a^{(2)}_{-1}\ket{0}_1\otimes\ket{0}_2+\frac{(1+q)(1-t)}{(1-q)(1+t)}a^{(2)}_{-1}a^{(2)}_{-1}\ket{0}_1\otimes\ket{0}_2+a^{(2)}_{-2}\ket{0}_1\otimes\ket{0}_2.\nn
\ea

The action of the R-matrix can be computed to 
\ba
&&{\bf R}(u)\ket{\tyng(1,1),\emptyset}=\frac{p(1+q)(1-t)(1-(1-p)(1+q)u)}{(1-q)(1+t)}a^{(1)}_{-1}a^{(1)}_{-1}\ket{0}_1\otimes\ket{0}_2\nn\\
&&+p(1-(1-p)(1+q)u)a^{(2)}_{-2}\ket{0}_1\otimes\ket{0}_2+\frac{(1-p)(1+q)(1-t)}{1+t}ua^{(2)}_{-1}a^{(2)}_{-1}\ket{0}_1\otimes\ket{0}_2\nn\\
&&+(1-p)(1+q)ua^{(2)}_{-2}\ket{0}_1\otimes\ket{0}_2+\frac{2p^{1/2}q(1-p)(1-t)(1+q)}{(1-q)(1+t)}ua^{(1)}_{-1}a^{(2)}_{-1}\ket{0}_1\otimes\ket{0}_2+\cO(u^2)\nn\\
&&=\frac{(u-1)(qu-1)}{(u-1/p)(qp u-1)}\ket{\emptyset,\tyng(1,1)}^{op}+\cO(u^2),\nn
\ea
\ba
&&{\bf R}(u)\ket{\tyng(2),\emptyset}=-\frac{p}{t}((1-p)(1+t)-t)ua^{(1)}_{-2}\ket{0}_1\otimes\ket{0}_2-\frac{u}{t}(1-p)(1+t)a^{(2)}_{-2}\ket{0}_1\otimes\ket{0}_2\nn\\
&&-\frac{p}{t}u((1-p)(1+t)-t)a^{(1)}_{-1}a^{(1)}_{-1}\ket{0}_1\otimes\ket{0}_2-\frac{u}{t}(1-p)(1-t)a^{(2)}_{-1}a^{(2)}_{-1}\ket{0}_1\otimes\ket{0}_2\nn\\
&&+\frac{2p^{1/2}}{t}(1-p)ua^{(1)}_{-1}a^{(2)}_{-1}\ket{0}_1\otimes\ket{0}_2+\cO(u^2)\nn\\
&&=\frac{(u-1)(u-t)}{(u-1/p)(pu-t)}\ket{\emptyset,\tyng(2)}^{op}+\cO(u^2),\nn
\ea
\ba
&&{\bf R}(u)\ket{\tyng(1),\tyng(1)}=\lt(p-\frac{1}{t}(1-q+qt-p+pq+pt-pqt-p^2t)u\rt)a^{(1)}_{-1}a^{(2)}_{-1}\ket{0}_1\otimes\ket{0}_2\nn\\
&&+\frac{(1-p)(1-q)(1+t)}{2p^{1/2}t}ua^{(2)}_{-2}\ket{0}_1\otimes\ket{0}_2+\frac{1}{sp^{1/2}t}(1-p)(1-(1-t)q-(1-2p)t)ua^{(2)}_{-1}a^{(2)}_{-1}\ket{0}_1\otimes\ket{0}_2\nn\\
&&+\cO(u^2)=\frac{(u-p)(u/p-q)}{(qu-1)(u-t)}\ket{\tyng(1),\tyng(1)}^{op}+\cO(u^2),\nn
\ea
\ba
&&{\bf R}(u)\ket{\emptyset,\tyng(2)}=\lt(p-(2-q+\frac{q}{t}-2p-\frac{q(1-t)}{pt})u\rt)a^{(2)}_{-1}a^{(2)}_{-1}\ket{0}_1\otimes\ket{0}_2\nn\\
&&-\lt(p-\frac{q(1-p)(1+t)}{pt}u\rt)a^{(2)}_{-2}\ket{0}_1\otimes\ket{0}_2-\frac{u}{t}(1-p)(q-pt)a^{(1)}_{-2}\ket{0}_1\otimes\ket{0}_2\nn\\
&&-\frac{2(1-p)(1-t)(q-pt)}{p^{1/2}t}ua^{(1)}_{-1}a^{(2)}_{-1}\ket{0}_1\otimes\ket{0}_2+\frac{(1-p)(1-t)(q-pt)}{t(1+t)}ua^{(1)}_{-1}a^{(1)}_{-1}\ket{0}_1\otimes\ket{0}_2+\cO(u^2)\nn\\
&&=\frac{(u-p)(tu/p-1)}{(u-1)(tu-1)}\ket{\tyng(2),\emptyset}^{op}+\cO(u^2),\nn
\ea
\ba
&&{\bf R}(u)\ket{\emptyset,\tyng(1,1)}=\lt(p-\frac{(1+q)(1-p)}{pt}u\rt)a^{(2)}_{-2}\ket{0}_1\otimes\ket{0}_2+\frac{u}{t}(1-p)(q-pt)a^{(1)}_{-2}\ket{0}_1\otimes\ket{0}_2\nn\\
&&-\frac{(1-p)(1-t)(q-pt)}{t(1+t)}ua^{(1)}_{-1}a^{(1)}_{-1}\ket{0}_1\otimes\ket{0}_2+\frac{2(1-p)(1+q)(1-t)(q-pt)}{p^{1/2}qt(1+t)}ua^{(1)}_{-1}a^{(2)}_{-1}\ket{0}_1\otimes\ket{0}_2\nn\\
&&-\frac{(1+q)(1-t)(p^2t+(2tp^2-(1-q)(1-p)-2tp)u)}{p(1-q)t(1+t)}a^{(2)}_{-1}a^{(2)}_{-1}\ket{0}_1\otimes\ket{0}_2+\cO(u^2)\nn\\
&&=\frac{(u-p)(u/p-q)}{(u-1)(u-q)}\ket{\tyng(1,1),\emptyset}^{op}+\cO(u^2).\nn
\ea

\bibliography{r-matrix_boson}
   
\end{document}